\documentclass[journal]{IEEEtran}
\ifCLASSINFOpdf
\usepackage[pdftex]{graphicx}
\usepackage{float}
\else
\fi
\begin{document}
%
\title{IMAGE COMPRESSION WITH LEARNED LIFTING-BASED DWT AND LEARNED TREE-BASED ENTROPY MODELS}
%
%
%

\author{Ugur Berk Sahin, 
        Fatih Kamisli
\thanks{Both authors are with the Department of Electrical and Electronics Engineering at the Middle East Technical University, Ankara, Turkey. (emails: u.berksahin@gmail.com, kamisli@metu.edu.tr)}
\thanks{Codes are available at https://github.com/uberkk/ImageCompression\\LearnedLiftingandLearnedTreeBasedModels}}

\maketitle

\begin{abstract}
This paper explores learned image compression based on traditional and learned discrete wavelet transform (DWT) architectures and learned entropy models for coding DWT subband coefficients. A learned DWT is obtained through the lifting scheme with learned nonlinear predict and update filters. Several learned entropy models are proposed to exploit inter and intra-DWT subband coefficient dependencies, akin to traditional EZW, SPIHT, or EBCOT algorithms. Experimental results show that when the proposed learned entropy models are combined with traditional wavelet filters, such as the CDF 9/7 filters, compression performance that far exceeds that of JPEG2000 can be achieved. When the learned entropy models are combined with the learned DWT, compression performance increases further. The computations in the learned DWT and all entropy models, except one, can be simply parallelized, and the systems provide practical encoding and decoding times on GPUs.  
\end{abstract}


%
\IEEEpeerreviewmaketitle

\section{Introduction} \label{sec:intro}
With the success of deep learning algorithms in computer vision applications, many image processing applications, including image compression, have been revisited with deep learning based algorithms \cite{jiao2019survey}. The compression performance of learning based image compression methods have improved quickly in recent years and have reached or exceeded those of state-of-the-art traditional image compression algorithms, such as JPEG2000 \cite{jp2k, JPEG2000} and HEIC \cite{HEVC} or AVIF \cite{avif}. 

Most of the recently proposed learning based image compression algorithms are based on the auto-encoder architecture \cite{Goodfellow-et-al-2016}. In this architecture, a convolutional neural network (CNN) is used to map the input image to a latent domain, which may have constraints on it such as a small entropy for compression applications. Another CNN is used to map the latent representation back to the image domain to approximate the input image. This idea is similar to the traditional transform coding approach \cite{goyal2001}, except that the analysis and synthesis functions are nonlinear, comprise CNN and are learned from real world images, whereas in traditional transform coding the analysis and synthesis functions are typically linear and derived based on simple models of images \cite{ahmed1974discrete, han2011jointly, kamisli2015block}. In most learning based compression systems, a CNN based system that learns the joint probability distribution of the quantized latent representation is combined with the auto-encoder architecture to obtain an end-to-end learned image compression system \cite{balle2016end, balle2018variational}. The overall system is trained to  minimize a rate-distortion cost function over a training set.

Traditional image compression systems are based on linear transforms, linear predictors and entropy models adapted to the specific systems. While JPEG, HEIC and AVIF are block-based systems, 
JPEG2000 is based on the Discrete Wavelet Transform (DWT), which processes entire images or subbands. 

DWT based systems, such as JPEG2000, process only the low-pass subband in next level wavelet transformations and produce a multi-resolution transform domain representation. This multi-resolution transform domain representation typically has dependencies in a spatial neighborhood within a subband and across the subbands in different DWT levels \cite{buccigrossi1999image, Zhen2002}. The dependency across subbands is exploited in compression applications with algorithms such as embedded zerotree wavelet (EZW) coding \cite{shapiro1993embedded} or set partitioning in hierarchical trees (SPIHT) \cite{said1996new} algorithms and the spatial dependency in a subband is exploited with algorithms such as embedded block coding with optimal truncation (EBCOT)\cite{taubman2000high}. 

This paper explores a learned image compression system that has an overall architecture which is inline with the traditional DWT based image compression systems but uses learned nonlinear filters to construct both a DWT and tree-based entropy models of the DWT subbands. In particular, the DWT is obtained through the lifting scheme with learned nonlinear predict and update filters, and several learned entropy models are proposed that exploit dependencies of DWT coefficients across subbands in different levels as well as within a subband, akin to traditional EZW, SPIHT or EBCOT algorithms. The resulting systems provide compression performance that far exceeds that of the traditional DWT based image compression system JPEG2000 at the expense of increased computational complexity but can still provide practical encoding and decoding times on GPUs. 

The remainder of the paper is organized as follows. Section \ref{sec:relwrk} reviews related background and recent research. Section \ref{ssec:lrnddwt} and \ref{ssec:lrndtree} present the proposed learning-based algorithms: the learned lifting-based DWT and several learned entropy models, respectively. Section \ref{sec:expres} presents experimental results and comparisons, and Section \ref{sec:con} concludes the paper.

\section{Related Work}\label{sec:relwrk}
This section reviews related background in wavelet transforms, coding and research in learned image compression.
\subsection{Lifting scheme and the Discrete Wavelet Transform (DWT)} \label{ssec:lif}
\begin{figure}[th]
\centerline{\includegraphics[trim=0 11 0 7, clip, width=\linewidth]{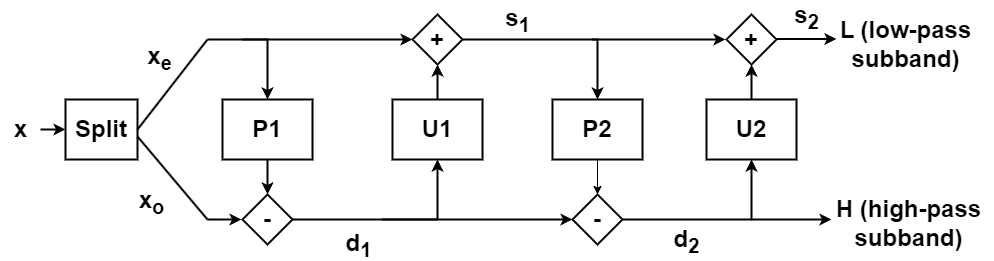}}
\caption{Lifting scheme.}
\label{fig:lif}
\end{figure}

The lifting scheme is a technique to obtain wavelet transforms \cite{sweldens1995lifting} and can also provide computation advantages for the wavelet transform implementation. Known wavelet transforms can also be factored into lifting steps \cite{daubechies1998factoring}. The lifting scheme can be summarized using the block diagram in Figure \ref{fig:lif}.

First, the signal $x$ is split into even and odd indexed samples $x_e$ and $x_o$, respectively. Next, odd indexed samples $x_o$ are predicted by processing even indexed samples $x_e$ using a predictor $P1(.)$. The prediction result is subtracted from $x_e$ to obtain the detail signal $d_1$, which is then processed with an update filter $U1(.)$ and added to the even indexed signal $x_e$ to obtain the smoothed signal $s_1$. Next, another round of prediction and update steps can be applied starting with $s_1$ and $d_1$ using $P2(.)$ and $U2()$ as shown in Figure \ref{fig:lif} to obtain detail and smoothed signals $d_2$ and $s_2$. In a similar manner, one can add more prediction and update steps. The final smoothed and detail signals, $s_2$ and $d_2$ in Figure \ref{fig:lif}, make up the low-pass and high-pass subbands of the wavelet transform. Scaling coefficients can also be used to scale the subbands but are not shown in Figure \ref{fig:lif} \cite{daubechies1998factoring}. Note that a lifting scheme based wavelet transform is always invertible no matter how the predict and update blocks are chosen. The inverse transform can be obtained by going in the reverse direction in the block diagram, i.e. starting with signals $s_2$ and $d_2$, and replacing additions with subtractions and vice versa to obtain back $x_e$ and $x_o$ and then the original signal $x$. 

If simple linear filters are used for $P(.)$ and $U(.)$ blocks, one obtains linear discrete wavelet transforms (DWT). For example, to obtain the DWT with the CDF 9/7 filters \cite{cohen1992biorthogonal} used in JEPG2000, one needs two $P(.)$ and $U(.)$ blocks (all with linear 2-tap filters) and scaling coefficients to scale $s_2$ and $d_2$ \cite{daubechies1998factoring}. We will propose to use nonlinear prediction and update blocks consisting of CNNs in Section \ref{ssec:lrnddwt} in order to improve the image compression performance with DWTs.

\begin{figure}[th]
\centerline{\includegraphics[trim=0 7 0 4, clip, width=\linewidth]{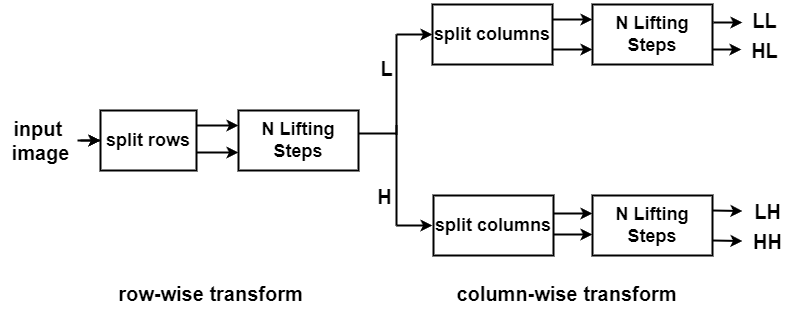}}
\caption{2-D Discrete Wavelet Transform (DWT) with lifting implementation}
\label{fig:2ddwt}
\end{figure}

For 2-D signals or images, one can apply the lifting scheme based DWT two times successively to obtain a 2-D DWT. First, one splits the 2-D signal into even and odd indexed rows and performs the first DWT along the columns. Next, the resulting two subbands are each split into even and odd indexed columns and undergo another lifting based DWT along the rows, resulting in a total of four subbands,  shown as LL, LH, HL, and HH (where L denotes low-pass and H denotes high-pass) in Figure \ref{fig:2ddwt}.

\begin{figure}[th]
\centering
\begin{minipage}{0.45\linewidth}
\centerline{\includegraphics[width=0.80\linewidth]{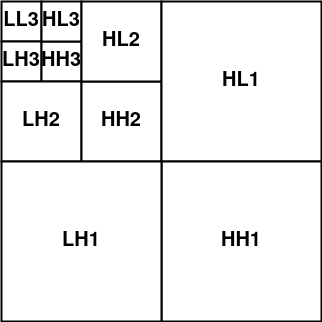}}
\caption{Visual illustration of subbands of 3-level 2-D DWT}
\label{fig:subb}
\end{minipage}
\begin{minipage}{0.45\linewidth}
\centerline{\includegraphics[width=0.80\linewidth]{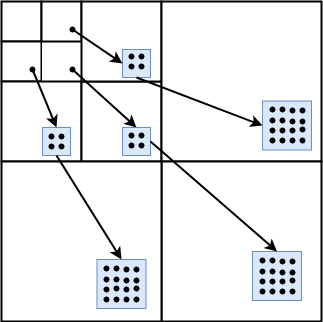}}
\caption{Visual illustration of zerotrees}
\label{fig:zerotree}
\end{minipage}
\end{figure}

To obtain a multi-level DWT, one can apply the lifting scheme repeatedly to the low-pass (smoothed) subband output of the previous level (in the case of 2-D DWT to the LL subband). This results in a multi-level DWT with multi-resolution subbands. In the case of 2-D DWT, the last level DWT will have four subbands (subbands ending with a 3 in Figure \ref{fig:subb}) and the previous levels will each have three subbands (subbands ending with a 2 or 1 in Figure \ref{fig:subb}). Notice that each subband has half the resolution (i.e. number of samples) of the subband in the previous level, along both dimensions in the case of 2-D DWT, leading to a multi-resolution transform domain representation.

\subsection{Wavelet coding with tree based structures} \label{ssec:tree}
It is well known that DWT coefficients coming from the same spatial neighborhood that are in different levels of a subband (e.g. HL1, HL2) are highly dependent \cite{dragotti2003wavelet, buccigrossi1999image, Zhen2002}. This property has been used in many image processing applications that use wavelet transforms, such as denoising \cite{dragotti2001footprints}, deconvolution \cite{dragotti2002deconvolution} or compression \cite{shapiro1993embedded, said1996new}.

In the embedded zerotree wavelet algorithm (EZW) \cite{shapiro1993embedded} and the set partitioning in hierarchical trees (SPIHT) algorithms \cite{said1996new}, this property is used in a tree-based structure. At low bit rates, i.e. high compression ratios, most of the DWT coefficients in high-pass subbands will be below a threshold, i.e. quantized to zero, since images contain many smooth regions. The locations of these zero coefficients, i.e. significance map, must be coded efficiently. Forming a tree structure by considering the DWT coefficients in high-pass subbands with the highest level subband coefficients at the root node and with the children of each tree node being the spatially related coefficients in the previous level subband, there is a high probability that one or more subtrees will consist entirely of coefficients which are zero or nearly zero. Such subtrees are called zerotrees (see Figure \ref{fig:zerotree}). Zerotrees allow efficient coding of many zeros in the significance map and thus enable successful compression performance.

EZW or SPIHT algorithms model the dependency across subbands in different levels using zerotrees or other possibly non-optimal algorithms. For optimal exploitation of the dependencies, the joint probability distribution of these wavelet coefficients is needed. In Section \ref{ssec:lrndtree}, we will propose to use the representation power of CNNs to model and learn the joint or conditional probabilities of these wavelet coefficients to improve the image compression performance with DWTs.

\subsection{Learned image compression} \label{ssec:lrndic}
Almost all the recently proposed learning based image compression algorithms are based on the auto-encoder architecture \cite{Goodfellow-et-al-2016}. In this architecture, the input image is transformed with a nonlinear analysis function that consists of a cascade of several (e.g. 4) convolutional neural network (CNN) layers and a nonlinear function in between them \cite{balle2016end}. The obtained latent domain representation has smaller spatial resolution (e.g. $\frac{1}{16}$ of the height and width of the original image) than the original image and has many channels or subbands (e.g. 128). The latent representation is scalar quantized and processed with a nonlinear synthesis function, that again consists of a cascade of several CNN layers and a nonlinear function in between them, to obtain the reconstructed image. An entropy model that learns the joint probability distribution of the quantized latent representation is combined with the auto-encoder architecture to obtain a learned image compression system \cite{balle2018variational, minnen2018joint}. The system is trained end-to-end to minimize a rate-distortion cost function over a training set. 

Numerous proposals appeared to improve the standard auto-encoder based compression system \cite{balle2016end, balle2018variational}. Some of these proposals focus on improving the auto-encoder \cite{cheng2020learned, yilmaz2021self, lu2021transformer}, but most focus on improving the entropy model \cite{balle2018variational,minnen2018joint,minnen2020channel,He_2022_CVPR, Kim_2022_CVPR} of the overall compression system.

\subsection{iWave and iWave++ algorithm} \label{ssec:iwave}
While most of the learning based image compression papers are based on auto-encoder architectures, there is one line of work, in addition to ours, that follows a learned DWT based compression architecture, named iWave \cite{ma2019iwave} and iWave++ \cite{ma2020end}. Ma et al. form a learned DWT by using the lifting scheme with CNN based predict and update blocks. They use two predict and update blocks for the wavelet transform in each DWT level and use 4 DWT levels in their experiments. 

While iWave \cite{ma2019iwave} uses the entropy coding algorithm in JPEG2000, the entropy model for the DWT subbands in iWave++ \cite{ma2020end} is a quite complicated deep learning-based design and also computationally demanding. This complicated entropy model consists of two parts. The first part is the long-term context extraction model, which contains 3 long short-term memory (LSTM) modules and takes as input the subbands one by one, starting with the highest level low-pass subband, and produces a long-term context that is used as an input to the second part of the model. The second part of the entropy model is a context fusion model, which takes as input the long-term context, produced by the long-term context extraction model, and combines it with the spatially causal context of a subband coefficient obtained through masked convolutions. The output are parameters  for the probability distribution model of the subband coefficients. 

This entropy model is quite complicated and also requires quite long encoding and decoding times in the order of twenty minutes for a single image from the Kodak dataset \cite{kodak}. Simplifications of neural network architectures of the entropy model and a parallelization approach to speed up the sequential processing (due to the masked convolutions in the context fusion model) were also proposed \cite{ma2020end}. These reduce the encoding and decoding times to twenty seconds with some reduction in compression performance. Ma et al. \cite{ma2020end} also use a post-processing module at the end of the system to improve the compression performance. Without the post-processing module, the performance drops significantly.


\section{Proposed Learned lifting based DWT} \label{ssec:lrnddwt}
This paper explores learned image compression based on a learned DWT and learned entropy models of the DWT subband coefficients. This section discusses the learned DWT while Section \ref{ssec:lrndtree} discusses the learned entropy models.

\begin{figure*}[t]
\centerline{\includegraphics[trim=0 2 0 1, clip, width=0.9\linewidth]{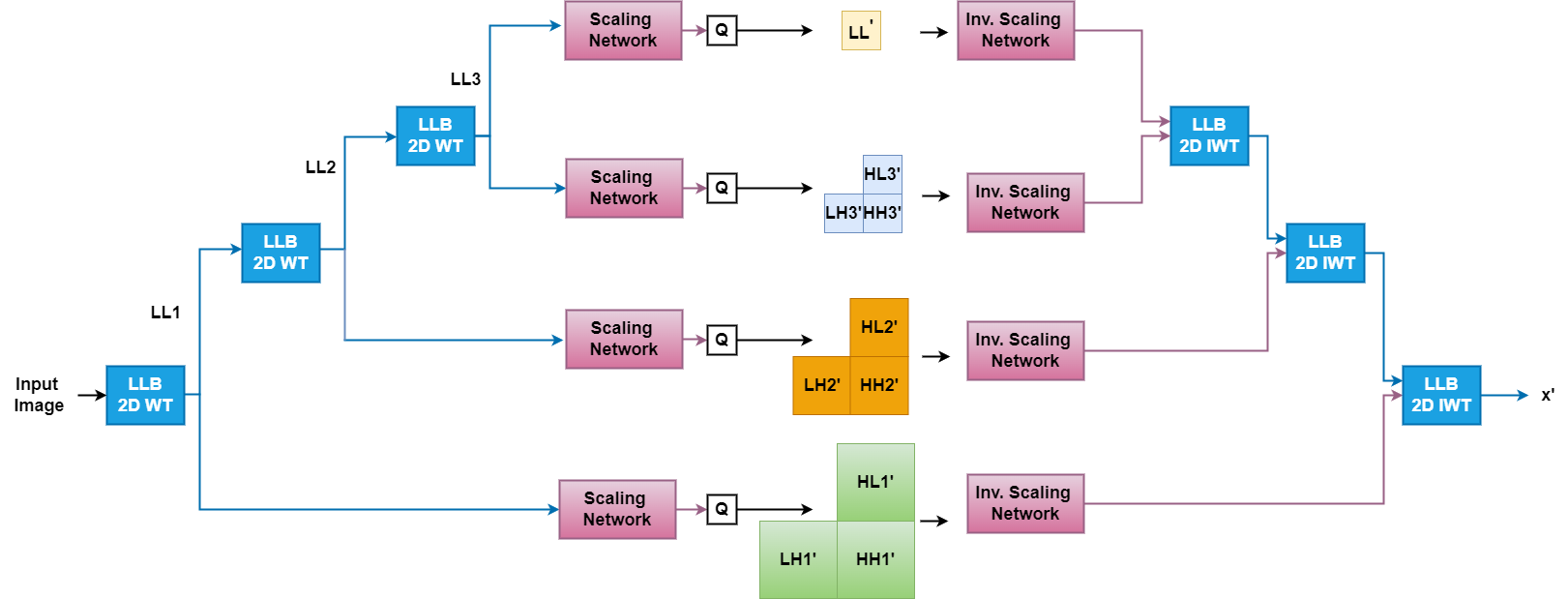}}
\caption{Overall learned lifting based DWT architecture (The figure shows a 3-level architecture while we use 4 levels in the experiments)}
\label{fig:dwtarch}
\end{figure*}

An overview of the learned lifting based DWT architecture is given in Figure \ref{fig:dwtarch}. We use a 4 level DWT architecture for our experiments but the figure shows a 3 level architecture for simpler illustration. The learned lifting based (LLB) 2-D wavelet transform (WT) in the figure is constructed with the lifting scheme comprising neural network based predict and update filters. The DWT subbands are processed with a scaling neural network to scale the subband coefficients prior to scalar quantization. The inverse transformation is also shown in the figure. The details of the learned wavelet transform and the scaling network are given below. 

\subsection{Learned lifting based (LLB) 2-D wavelet transform (WT) with learned predict and update filters} \label{sssec:lrndlif}
\begin{figure}[h]
\centerline{\includegraphics[trim=0 10 0 10, clip, width=0.27\linewidth]{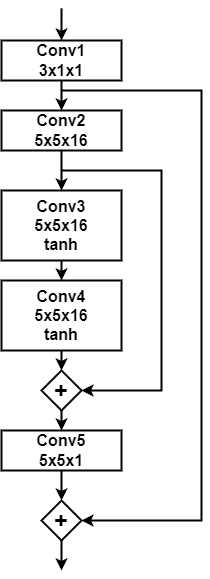}} 
\caption{Convolutional neural network (CNN) architecture used as predict ($P(.)$) and update ($U(.)$) filters}
\label{fig:P_U_filters}
\end{figure}

As discussed in Section \ref{ssec:lif}, a wavelet transform can be obtained through the lifting scheme. Instead of using simple linear filters for the predict and update blocks of the lifting scheme, convolutional neural networks (CNN) are used in this paper. We expected that the configuration of the CNN for the predict and update blocks not to be significantly important for the compression performance of the overall system, however, we found that this is not the case. In particular, we initially used a simple CNN architecture consisting of a few cascaded convolutional layers but obtained much better compression performance when we used the neural network architecture in Figure \ref{fig:P_U_filters}, which were used in \cite{ma2020end}. 

We attribute the good compression performance we obtained with the CNN architecture in Figure \ref{fig:P_U_filters} to the following two reasons, which are not explicitly discussed in \cite{ma2020end}. First, the CNN configuration in Figure \ref{fig:P_U_filters} has short-cut connections, which are known to be useful for good gradient flow in deep neural networks \cite{he2016deep}. Considering that we use two predict and two update filters for each wavelet transform (as in Figure \ref{fig:lif}) and build a DWT with 4 transform levels, the entire path from the input image to the highest level DWT subbands contain a total of 16 of the CNN architecture in Figure \ref{fig:P_U_filters}. Considering also that there are 5 convolutional layers in this CNN architecture, there are a total of 80 convolutional layers along that path. The overall neural network along this path is quite deep and short-cut or residual connections are useful for gradient flow and good learning performance \cite{he2016deep}.

Second, the weights of the first convolutional layer, Conv1, are initialized with the weights obtained by factorizing the well-known CDF 9/7 wavelet filters, that are used in JPEG2000, into the lifting scheme \cite{daubechies1998factoring}. In other words, if only the Conv1 layer is kept and all other Conv layers in Figure \ref{fig:P_U_filters} are removed (or replaced with zero weights), and two such predict and update filters are cascaded, then one obtains the CDF 9/7 wavelet transform. The benefit of having such a configuration in Figure \ref{fig:P_U_filters} is that the obtained wavelet transform is very likely to improve upon that of the CDF 9/7 filter as the first Conv1 layer implements the predict or update of the CDF 9/7 filter and the remaining Conv layers calculate an improvement that is added, due to the shortcut connection, on top of the CDF 9/7 predict or update result. An additional benefit is that the training becomes much more stable.

For the learned DWT architecture in this paper, the learned lifting scheme is applied with two predict and two update filters, first along the vertical dimension, and then along the horizontal dimension (with different parameters for the predict and update blocks) to obtain a learned lifting based (LLB) 2-D wavelet transform (WT). This learned WT is repeated on the low-pass subband 4 times (with the same CNN parameters) to obtain the overall learned DWT architecture shown in Figure \ref{fig:dwtarch}. Note also that the input image is converted to the YCbCr color space and the DWT is applied separately on each color channel (with different learned CNN parameters).

\subsection{DWT coefficient scaling network} \label{sssec:sclntw}
\begin{figure}[h]
\centering
\begin{minipage}{0.48\linewidth}
\centerline{\includegraphics[trim=0 10 0 11, clip, width=0.36\linewidth]{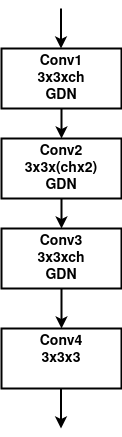}}
\centerline{\footnotesize (a)}
\end{minipage}
\begin{minipage}{0.48\linewidth}
\centerline{\includegraphics[trim=0 10 0 11, clip, width=0.36\linewidth]{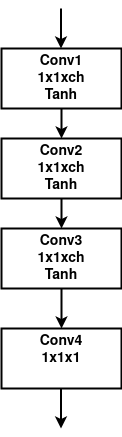}}
\centerline{\footnotesize (b)}
\end{minipage}
\caption{Scaling network used at the encoder with (a) the learned lifting based DWT (b) the traditional CDF 9/7 filters 
}
\label{fig:sclntw2}
\end{figure}

A scaling network is used both at the encoder, to process the subbands produced by the learned DWT, and at the decoder to process the received quantized subbands prior to the learned inverse DWT. The used scaling network at the encoder is shown in Figure \ref{fig:sclntw2} a). The decoder side scaling network is similar except that inverse GDN \cite{balle2018efficient} is used. The motivation behind using the scaling networks is multiple fold. 

First, quantization in our system is performed always with a unit quantization step-size no matter what the target bit-rate is (as in most learned compression systems in the literature \cite{balle2018variational}) and therefore the scaling network performs a scaling to the subband coefficients, prior to quantization, to adjust the range of the values of the subband coefficients. Second, it is known that dead-zone quantization improves compression performance of DWT based systems and is used in JPEG2000 \cite{marcellin2002overview}. The scaling network can also learn to scale the subband coefficient values non-linearly to have non-uniform quantization similar to dead-zone quantization. Third, convolution layers stacked with GDN layers are known to gaussianize image data \cite{balle2015density} and the subband coefficient distributions can better match a Gaussian distribution, which we  use in the entropy model as discussed in Section \ref{ssec:lrndtree}. 

Note that we are also interested in exploring how the entropy models we propose in Section \ref{ssec:lrndtree} work with traditional wavelet filters, in particular the CDF 9/7 wavelet filters \cite{cohen1992biorthogonal} used in JPEG2000. We again use a scaling network with the DWT that uses these CDF 9/7 filters but a much simpler one shown in Figure \ref{fig:sclntw2} b). The primary goal is to achieve the first two objectives of the scaling network discussed above. In summary, this simpler scaling network applies a scalar nonlinear mapping to each subband coefficient, which is different for each subband. 

Finally, note that the quantization operation (shown with Q in Figure \ref{fig:dwtarch}) on the scaled subband coefficients is replaced with additive independent noise uniform in $[-\frac{1}{2}, \frac{1}{2}]$ during training as the quantization operation kills gradients necessary to train the preceding scaling and LLB WT networks \cite{balle2016end, balle2018variational}. During compression (inference), quantization is performed.


\section{Proposed learned entropy models with tree-based structures} \label{ssec:lrndtree}
Several learned entropy models to code the quantized DWT subband coefficients are explored in this section. While our learned DWT is similar to that of \cite{ma2020end}, our learned entropy models are quite different. While \cite{ma2020end} use a complicated entropy model based on autoregressive models and LSTM modules, we explore several entropy models inspired by traditional wavelet coding approaches EZW, SPIHT and EBCOT to exploit the dependencies of DWT coefficients across subbands in different levels and within the same subband.

This section presents and discusses 4 entropy models that can be used with traditional linear wavelet transform filters or with the learned lifting based DWT architecture discussed above in Section \ref{ssec:lrnddwt}. These entropy models form different models for the joint probability distribution of the DWT subbands. The joint probability distribution of all subband coefficients are factored into a product of marginal and/or conditional distributions of each subband coefficient based on the chain rule of probability. The conditioning is performed differently in each of the entropy models based on exploitation of the inter subband (across subbands in different DWT levels) and/or intra subband (i.e. within a subband) dependencies of the subband coefficients.

\subsection{Factorized entropy model (FEM)} \label{sssec:fem}
The factorized entropy model (FEM) assumes that all subband coefficients are independent of each other, and thus uses marginal distributions for all subband coefficients to factor the joint distribution. It also assumes that all coefficients within the same subband are identically distributed. Thus, the number of probability distributions that are learned is simply equal to the number of subbands.

To learn each distribution, the neural network model in \cite{balle2018variational} with the implementation in the CompressAI library \cite{begaint2020compressai} is used. The approach in \cite{balle2018variational} obtains the probability values of quantized coefficients from the continuous (un-quantized) subband coefficient's cumulative distribution function (CDF), which is modeled using a fully connected neural network with some constraints to satisfy two required properties of a CDF: the range of the function must be the interval $[0, 1]$, and the function must be monotonically non-decreasing. These properties are imposed by having a logistic (sigmoid) function at the end of the neural network, and imposing some constraints on the weights of the neural network \cite{balle2018variational, chilinski2020neural}, respectively. 

\subsection{Tree-based inter subband conditioned entropy model (ISCEM)} \label{sssec:iscem}
\begin{figure}[h]
\centerline{\includegraphics[trim=0 5 0 14, clip, width=0.16\linewidth]{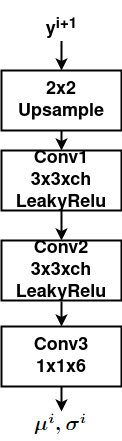}}
\caption{Convolutional neural network for inter subband conditioned probability model (ISCEM)
}
\label{fig:iscem}
\end{figure}

The inter subband conditioned entropy model (ISCEM) exploits the dependency of coefficients across subbands similar to EZW or SPIHT algorithms \cite{shapiro1993embedded, said1996new}. To exploit the inter subband dependency, the probability of a coefficient in a subband is conditioned on the coefficients in a higher subband and within the co-located spatial neighbourhood. This conditional probability is assumed to follow a Gaussian distribution, and thus only the mean and standard deviation $(\mu, \sigma)$ are needed. In particular, the neural network in Figure \ref{fig:iscem} is used to obtain the $(\mu, \sigma)$ parameters of each coefficient's probability model in a subband $y^i$ (e.g. LH2 in Figure \ref{fig:dwtarch}), conditioned on the co-located coefficients in a higher subband $y^{i+1}$ (e.g. LH3 in Figure \ref{fig:dwtarch}). The neural network performs zero-order-hold (ZOH) upsampling of subband $y^{i+1}$ and then processes it with two convolutional layers to produce the $(\mu^i, \sigma^i)$ parameter map for each coefficient in the subband $y^i$.

Notice that this entropy model forms a tree based dependency structure, similar to Figure \ref{fig:zerotree} of EZW or SPIHT algorithms, since the receptive field of the CNN in Figure \ref{fig:iscem} is confined to the co-located spatial neighborhood in the higher level subband. However, the dependency across subbands is exploited with probability distributions via neural network based models instead of the more traditional and handcrafted algorithms in EZW or SPIHT.

All subbands, except the highest level subbands are coded with the discussed ISCEM. The highest level subbands (i.e. LL, LH3, HL3, HH3 in Figure \ref{fig:dwtarch}) have no subbands to condition on and thus are coded simply with the FEM.

\subsection{Inter and intra subband conditioned entropy model (IISCEM)} \label{sssec:iiscem}
\begin{figure}[t]
\centering
\begin{minipage}{0.50\linewidth}
\centerline{\includegraphics[trim=0 18 0 20, clip, width=0.65\linewidth]{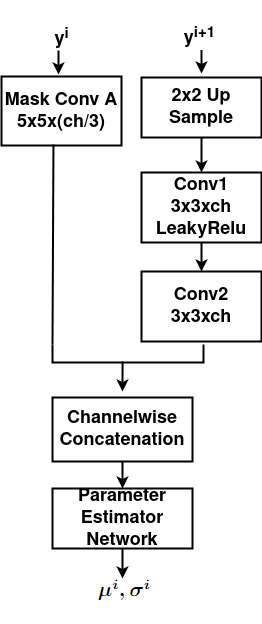}}
\caption{Convolutional neural network for inter and intra subband conditioned probability model (IISCEM) 
}
\label{fig:iiscem}
\end{minipage}
\hfill
\begin{minipage}{0.44\linewidth}
\centerline{\includegraphics[trim=0 0 0 0, clip, width=0.40\linewidth]{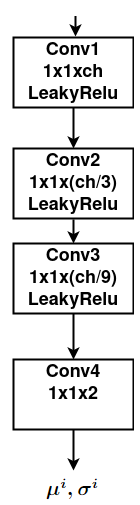}}
\caption{Parameter estimator network for the Gaussian distribution 
}
\label{fig:pen}
\end{minipage}
\end{figure}

The inter and intra subband conditioned entropy model (IISCEM) is similar to the above discussed ISCEM but also considers the intra subband dependency, i.e. dependency between spatially neighbor coefficients within a subband. The used convolutional neural network is shown in Figure \ref{fig:iiscem}. 

To exploit the inter subband dependency across subbands, the right branch of the CNN in Figure \ref{fig:iiscem} processes the higher subband $y^{i+1}$ with a network similar to that of ISCEM in Figure \ref{fig:iscem}. The left branch in Figure \ref{fig:iiscem} is responsible from exploiting the intra subband dependency, i.e.  dependency between spatially neighbor subband coefficients within a subband, and uses masked convolutions for this purpose. A simple and widely used method to exploit dependency between spatially neighbor coefficients is to process coefficients in raster scan order and use the previously obtained left and upper coefficients to process the next coefficient. To simulate such causal processing during training of deep neural networks, a common method is to use masked convolutions \cite{van2016conditional}, where the filter weights (of a regular convolution filter) that will multiply coefficients that are not to the left and top are  set to zero.

The result of the right and left branch in Figure \ref{fig:iiscem} are combined, by channel-wise concatenation, and further processed by a parameter estimator network, shown in Figure \ref{fig:pen}, to produce the $(\mu^i, \sigma^i)$ parameter map for each coefficient in the subband $y^i$, whose probability distribution is again modeled with a Gaussian distribution.


Note that the raster scan based causal processing in the left branch of Figure \ref{fig:iiscem} requires sequential processing during decoding because a subband coefficient has to be decoded before it can be used for the processing of the next subband coefficient. To decode a subband coefficient, the left branch has to perform processing (using only the relevant coefficients to the left and top, i.e. on a small patch) starting with the Mask Conv A layer until to the end of the Parameter Estimator Network. To decode all subband coefficients, the left branch has to perform processing sequentially as many times as the number of coefficients in subband $y^i$, while the right branch (up to Conv2 output) has to perform processing only once with the entire subband $y^{i+1}$ as input. The sequential processing required on the left branch during decoding does not allow for exploiting the massively parallel computation capabilities of GPUs, and therefore slows down the encoding and decoding speed significantly. Nevertheless, such models are used in many learned compression models due to the provided improvement in compression performance \cite{minnen2018joint, cheng2020learned}.

\begin{figure}[h]
\centerline{\includegraphics[trim=0 6 0 7, clip, width=0.16\linewidth]{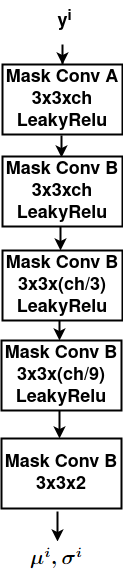}}
\caption{Convolutional neural network for highest level subbands in IISCEM. (Type A masked convolutions set the current pixel to zero, while type B do not \cite{salimans2017pixelcnn++}, to preserve causality.) 
}
\label{fig:iiscem_ll}
\end{figure}

All subbands, except the highest level subbands are coded with the discussed IISCEM. The highest level subbands (i.e. LL, LH3, HL3, HH3 in Figure \ref{fig:dwtarch}) have no subbands to condition on and thus need to be coded with a different entropy model. In order to obtain good compression performance, we exploit the spatial dependencies in these subbands and use a neural network architecture again based on masked convolutions, which is shown in Figure \ref{fig:iiscem_ll}.

Note that the encoding/decoding times with the IISCEM are orders of magnitude longer (as we show in Section \ref{sec:expres}) than with the other 3 entropy models we present in this section. Thus, IISCEM seems infeasible for practical use. Nevertheless, we included IISCEM here primarily to see how well it performs in comparison to the other entropy models.


\subsection{Inter and parallel intra subband conditioned entropy model (IPISCEM)} \label{sssec:ipiscem}
\begin{figure}[h]
\centerline{\includegraphics[trim=0 14 0 8, clip, width=0.95\linewidth]{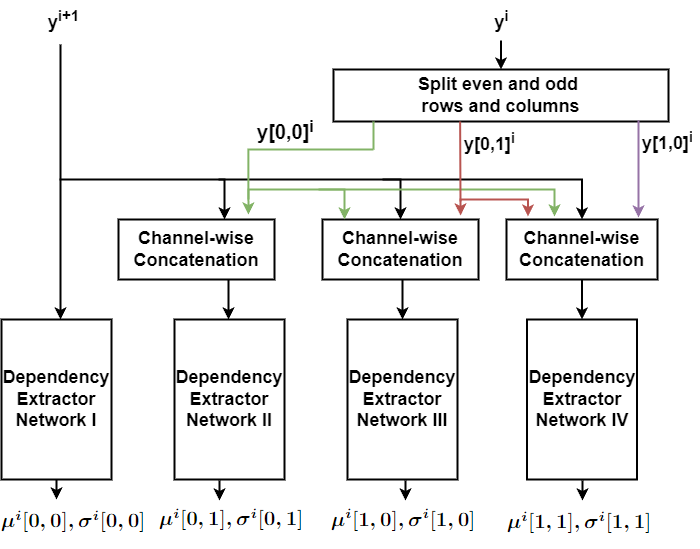}}
\caption{Convolutional neural network for inter and parallel intra subband conditioned probability model (IPISCEM)} 
\label{fig:ipiscem}
\end{figure}

\begin{figure}[h]
\centerline{\includegraphics[trim=0 10 0 8, clip, width=0.46\linewidth]{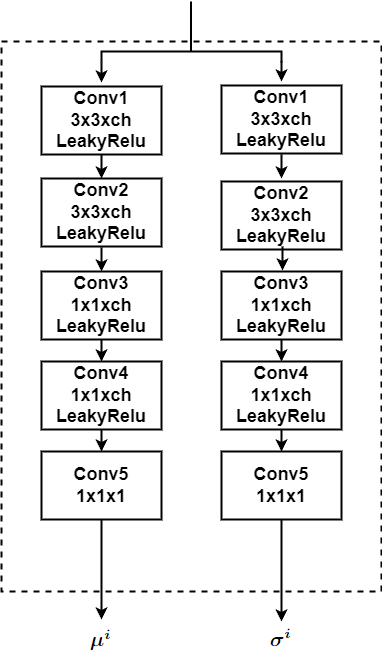}}
\caption{Dependency extractor network Parameter estimator network for the Gaussian distribution 
}
\label{fig:depen}
\end{figure}

To overcome the significant encoding/decoding speed drawback of IISCEM, an alternative model, the inter and parallel intra subband conditioned entropy model (IPISCEM), is proposed in this subsection. The IPISCEM model is shown in Figure \ref{fig:ipiscem}. The subband $y^i$, for which the probabilities are to be calculated, is divided into four  by splitting it into even and odd rows and columns denoted $y^i[0,0]$, $y^i[0,1]$, $y^i[1,0]$ and $y^i[1,1]$, respectively. The assumed Gaussian distribution parameters ($\mu^i[0,0]$, $\sigma^i[0,0]$) of $y^i[0,0]$ are obtained conditioned on the higher subband $y^{i+1}$ by processing $y^{i+1}$ with the Dependency Extractor Network I is shown in Figure \ref{fig:ipiscem}. Next, the probability distribution parameters ($\mu^i[0,1]$, $\sigma^i[0,1]$) of $y^i[0,1]$ are obtained conditioned on the higher subband $y^{i+1}$ and the previously obtained $y^i[0,0]$ by combining them, via channel-wise concatenation, and by processing them with the Dependency Extractor Network II. This procedure is repeated to obtain the probability distribution parameters ($\mu^i[1,0]$, $\sigma^i[1,0]$) and ($\mu^i[1,1]$, $\sigma^i[1,1]$) of $y^i[1,0]$ and $y^i[1,1]$, respectively, as shown in Figure \ref{fig:ipiscem}. The neural network model of the dependency extractor networks is shown in Figure \ref{fig:depen}.

The IPISCEM model exploits both the inter and intra subband dependency of subbands in 4 sequential steps, which all contain CNN that can take advantage of parallel computation capabilities of GPUs. Hence, IPISCEM does not suffer from the decoding speed drawback of IISCEM but still exploits  intra subband dependency for improved compression performance.  

All subbands, except the highest level subbands are coded with the discussed IPISCEM. The highest level subbands (i.e. LL, LH3, HL3, HH3 in Figure \ref{fig:dwtarch}) have no subbands to condition on and thus are coded simply with the FEM. Note that while a masked convolution based architecture as in IISCEM could be used for improved compression performance  (see Figure \ref{fig:iiscem_ll}), we simply used the FEM to avoid slow encoding/decoding.

\section{Experiments and Results}\label{sec:expres}

\subsection{Experimental settings} \label{ssec:expset}
The Vimeo Triplet dataset \cite{vimeo} is used for all training performed for this paper. Randomly cropped patches of size 256x256 were used from the training set with a batch-size of 8. The Adam optimizer with an initial learning rate of $10^{-4}$ was used together with a learning rate scheduler that reduces the learning rate when the cost plateaus. All implementation for the proposed systems in this paper were done with PyTorch 
and the CompressAI library  \cite{begaint2020compressai}. Our codes are shared on github at the link in \cite{berkgithub}.

\subsection{Compression results} \label{ssec:compres}
The compression tests were conducted using the Kodak image dataset \cite{kodak} that contains 24 RGB images of 512x768 resolution. All systems are trained multiple times for different rate-distortion trade-offs by changing the $\lambda$ weight in the rate-distortion cost ($R+\lambda D$). The range asymmetric numeral system (rANS) implementation in \cite{begaint2020compressai} is used as an entropy coder to generate the compressed bitstreams.

\subsubsection{Learned entropy models with CDF 9/7 filters} \label{sssec:resent}
\begin{figure}[t]
\centerline{\includegraphics[trim=25 8 32 25,clip,width=0.90\linewidth]{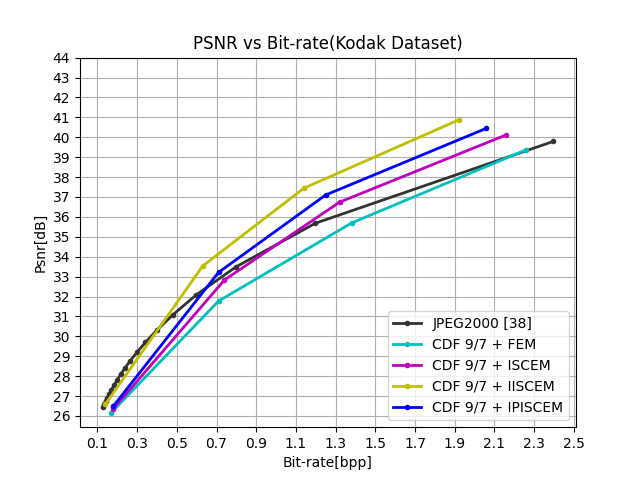}}
\caption{Compression performance results of learned tree-based entropy models with standard CDF 9/7 wavelet filters on the Kodak dataset}
\label{fig:rescdf97}
\end{figure}

First, we are interested in the performance of the entropy models with standard linear wavelet filters used in JPEG2000, i.e. the CDF 9/7 wavelet filters \cite{cohen1992biorthogonal}. Thus, we form a transform architecture that performs a 4-level DWT transform with the CDF 9/7 wavelet filters by replacing the learned lifting based wavelet transforms (LLB WT) in Figure \ref{fig:dwtarch} with the CDF 9/7 wavelet filters and use the simple scaling network in Figure \ref{fig:sclntw2} b). The results are shown in Figure \ref{fig:rescdf97}.

As expected, the system based on FEM has the worst compression performance, as it does not exploit any dependency of the subband coefficients. The system based on ISCEM has better performance as it exploits inter subband dependency, i.e. dependency of subband coefficients across successive subbands in the same spatial neighborhood. The system based on IISCEM has even better performance as it exploits also the intra subband dependency in addition to inter subband dependency. The system based on IPISCEM, which includes an alternative parallel implementation friendly method to exploit intra subband dependency, has lower performance than IISCEM but higher performance than ISCEM.

When the performance of these four systems are compared with that of JPEG2000, it can be seen that all systems, except the one based on IISCEM, start off with lower performance at the low bitrate end but catch up or are significantly superior at higher bitrates. At the very low bitrate end, most subband coefficients except the ones in the highest level subbands (e.g. LL, HL3, LH3, HH3 in Figure \ref{fig:dwtarch}) are mostly quantized to zero, and thus efficient coding of the highest level subbands becomes very important. Both JPEG2000 and IISCEM exploit the intra subband dependency in these highest level subbands while the FEM, ISCEM and IPISCEM use the factorized entropy model primarily to achieve fast encoding and decoding speeds. At mid to high bitrates, ISCEM, IISCEM and IPISCEM provide significantly better compression than the entropy model in JPEG2000, indicating that the proposed learned entropy models can provide better compression than traditional entropy models.

Table \ref{table:param_number} shows the encoding and decoding speeds of the explored systems\footnote{Encoding/decoding times for JPEG2000, and the two systems with IISCEM are obtained with a CPU (due to sequential encoding/decoding requirement) while all others are obtained with a GPU}. Consider the first four rows (remaining rows will be discussed below). All systems except the one based on IISCEM have encoding and decoding times below one second. IISCEM has much higher encoding and decoding times since it uses an autoregressive model to exploit intra subband dependency, which requires sequential processing of subband coefficients during encoding and decoding, as discussed in Section \ref{sssec:iiscem}. Hence, while IISCEM provides better compression than IPISCEM, IPISCEM provides a more practical method to exploit intra subband dependency.

Overall, one general conclusion that can be drawn from these results is that it is possible to obtain better compression with learned entropy models than with traditional systems for wavelet based image coding, though with higher computational complexity but the computations can be parallelized.

\begin{table}[t]
\centering
\caption{Comparison of number of parameters and encoding/decoding times for a single 512x768 image from Kodak dataset}
\label{table:param_number}
\begin{tabular}{l|r|r|r}
\hline
\textbf{Compression system} & \textbf{Parameters}        &  \textbf{Encoding}   & \textbf{Decoding }   \\ 
                            &                            &  \textbf{time(sec)}  & \textbf{time(sec)}   \\ \hline
CDF 9/7      + FEM          &  0.17 M                    & 0.09    & 0.07 \\ \hline
CDF 9/7      + ISCEM        &  4.80 M                    & 0.22    & 0.17  \\ \hline
CDF 9/7      + IISCEM (cpu) &  7.70 M                    & 375.47  & 1277.68     \\ \hline
CDF 9/7      + IPISCEM      &  2.67 M                    & 0.62    & 0.57  \\ \hline
LLB DWT      + FEM          &  9.40 M                   & 0.28    & 0.23  \\ \hline
LLB DWT      + ISCEM        & 14.04 M                   & 0.41    & 0.33 \\ \hline
LLB DWT      + IISCEM (cpu) & 16.94 M                   & 377.78  & 1280.88 \\ \hline
LLB DWT      + IPISCEM      & 11.91 M                   & 0.80    & 0.72 \\ \hline
iWave++ w/ Post-pr(Slow)\cite{ma2020end}  & 17.91 M    & 1128.17 & 1153.46 \\ \hline
iWave++ w/ Post-pr(Fast)\cite{ma2020end}  &  1.29 M    &   22.17 &   24.82 \\ \hline
JPEG2000 (cpu)              & --                         &    0.14 &    0.08 \\ \hline
\end{tabular}
\end{table}

\subsubsection{Learned lifting based (LLB) DWT and learned entropy models} \label{sssec:resllb}
\begin{figure}[t]
\centerline{\includegraphics[trim=25 8 32 25,clip,width=0.9\linewidth]{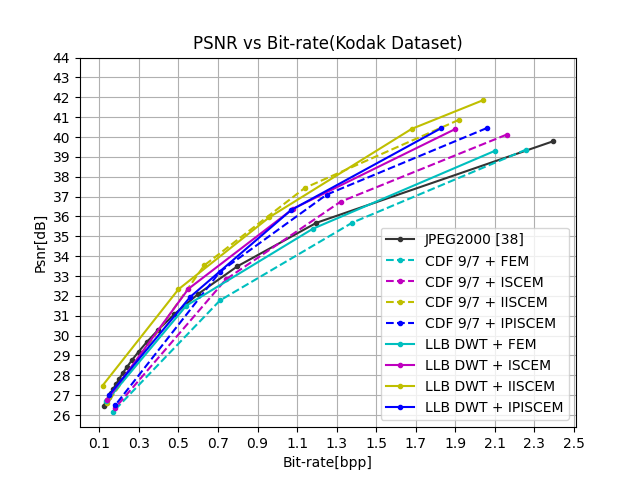}}
\caption{Compression performance results of learned tree-based entropy models with learned lifting based DWT on the Kodak dataset}
\label{fig:resllb}
\end{figure}

Next, we use the learned lifting based (LLB) DWT and the more complex scaling network together with the learned entropy models to obtain further improved compression performances. The compression results are shown in Figure \ref{fig:resllb}. As expected, the compression performance improves with every entropy model when LLB WT are used instead of CDF 9/7 filters. 

Table \ref{table:param_number} shows that the encoding and decoding times do not change significantly when LLB DWT is used instead of CDF 9/7 filters, which shows that the entropy modeling networks are the main components that take time in our proposed architectures.

\begin{figure}[t]
\centerline{\includegraphics[trim=25 8 32 25,clip,width=0.9\linewidth]{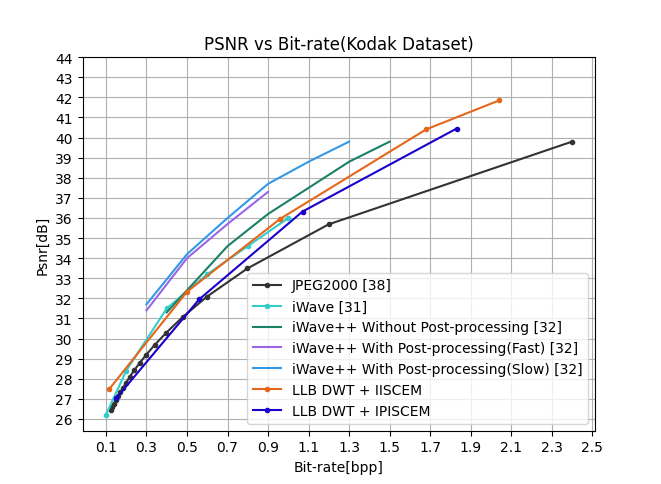}}
\caption{Compression performance comparisons of learned tree-based entropy models with learned lifting based DWT and other methods in the literature}
\label{fig:resall}
\end{figure}

We also compare the obtained systems based on LLB DWT and IISCEM/IPISCEM with other learned image compression methods in the literature, in particular with iWave \cite{ma2019iwave} and iWave++ \cite{ma2020end} as they are the only DWT based learned systems. Compression performance comparisons can be seen in Figure \ref{fig:resall}. The iWave++ with post-processing module \cite{ma2020end} system has the best performance since it has a LSTM module and an autoregressive component in the entropy model, but these components also lead to very long encoding/decoding times (1128.17/1153.46 seconds) with a GPU (RTX 2080 Ti), as shown in Table \ref{table:param_number}. In addition, a post-processing module is used, which is removed causes a big drop in the compression performance as seen in Figure \ref{fig:resall}. We also experimented with post-processing modules for our systems but could not get any significant compression gains as those in iWave++. The authors of iWave++ also propose a fast model by simplifying their entropy model, which causes a drop in compression performance as seen in Figure \ref{fig:resall}, but the encoding/decoding times are still quite long (22.17/24.82 seconds) as seen in Table \ref{table:param_number}, much longer than in our systems.   

Overall, our systems, such as LLB DWT + ISCEM or IPISCEM, have worse compression performance than iWave++ at mid-to-low bitrates but seem to catch up at higher bitrates and provide much smaller encoding decoding times (fraction of a second) as can be inspected from Table \ref{table:param_number} and thus are much more suitable for practical use. Compared to JPEG2000, our systems can achieve much higher compression performance, especially at high bitrates while still providing small encoding and decoding times (except for IISCEM) with GPUs.

\subsection{Visual results} \label{ssec:visres} 
\begin{figure*}[t]
\centering
\includegraphics[trim=90 130 110 230, clip, width=0.24\linewidth]{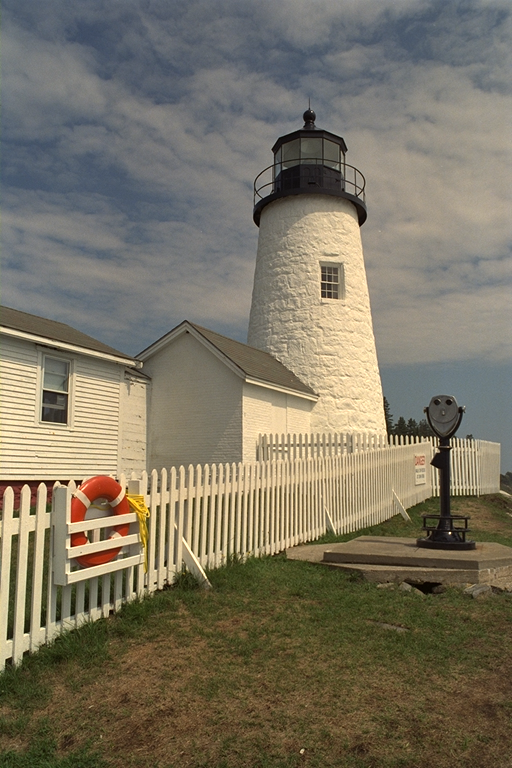} 
\includegraphics[trim=90 130 110 230, clip, width=0.24\linewidth]{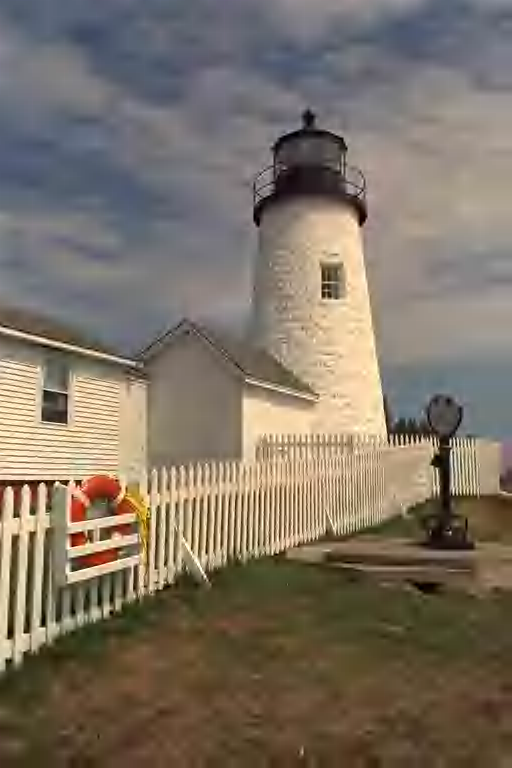}
\includegraphics[trim=65  93  81 169, clip, width=0.24\linewidth]{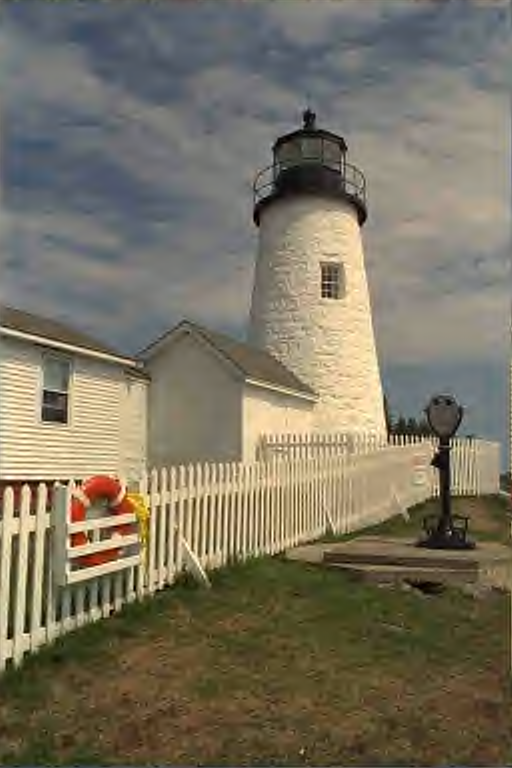}
\includegraphics[trim=65  93  81 169, clip, width=0.24\linewidth]{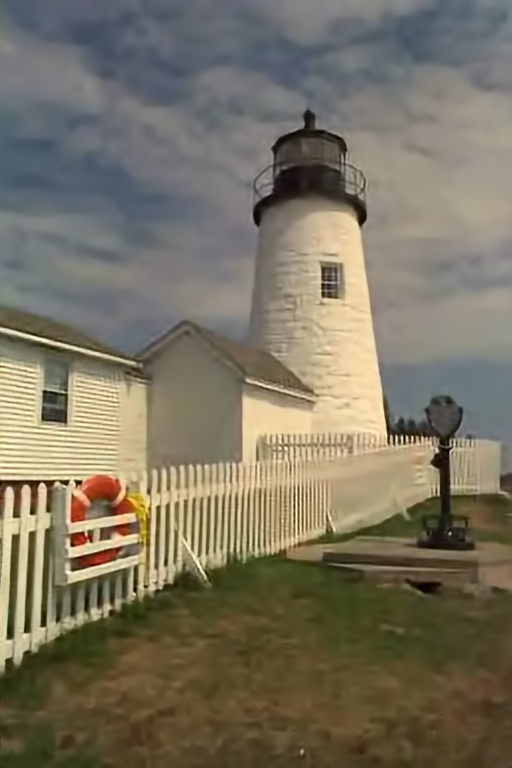} 
\caption{Comparison of visual properties of reconstructed pictures from left to right with (PSNR, bpp): Original, JPEG2000 (26.50, 0.120), CDF 9/7 + IPISCEM (26.16, 0.186), LLB DWT + IPISCEM (26.50, 0.146). }
\label{fig:recon_imgs}
\end{figure*}

Figure \ref{fig:recon_imgs} shows an original image and its compressed version by JPEG2000, CDF 9/7 + IPISCEM and LLB DWT + IPISCEM systems around similar picture quality of $26.50$dB PSNR. It can be seen that the picture reconstructed by JPEG2000 has ringing artifacts around the edges of the roof and is quite blurry in the grass region. The picture reconstructed by CDF9/7 + IPISCEM also has ringing artifacts but the blurriness is reduced. Ringing artifacts are expected in this system, too, since the same CDF 9/7 filters are used. On the other hand, the picture reconstructed by LLB DWT + IPISCEM has almost no ringing artifacts around edges, which are reconstructed much cleaner. This is similar to reconstructions of other learning based systems \cite{balle2016end}, which also reconstruct clean edges. Similar observations have also been made for other images and systems with CDF 9/7 or LLB DWT filters.

\section{Summary and Conclusions} \label{sec:con}
This paper explored learned image compression based on several learned entropy models for DWT based transform representations. These learned entropy models are designed to exploit the inter and intra subband dependency of wavelet coefficients. These entropy models can be used together with traditional linear wavelet filters, such the CDF 9/7 filters, or learned nonlinear wavelet filters, such as the one we used in this paper, to further improve the compression performance of the overall system. The resulting systems provide compression performance that far exceeds that of the traditional DWT based image compression system JPEG2000 at the expense of increased computational complexity, but the computations can be simply parallelized on GPUs and can provide practical encoding and decoding times.

\appendices
\section{}
Following parameters are used in experimental results. In Figure \ref{fig:sclntw2} a), $ch=32$ is used for processing LL and $ch=96$ is used for processing LH, HL, HH subbands together. In Figure \ref{fig:sclntw2} b), $ch=32$ is used for each subband. In Figure \ref{fig:iscem}, $ch=243$ is used for jointly processing LH, HL, HH subbands and output gives mean and scale for the corresponding three channels. In Figure \ref{fig:iiscem}, on the right hand side $ch=243$ is used for jointly processing LH, HL, HH subbands and $ch/3$ is used on the left handside for processing each subband LH,HL,HH separately (total of $243$ channels). In Figure \ref{fig:pen}, $ch=162$ is used for processing each LH, HL, HH subband separately. In Figure \ref{fig:iiscem_ll}, $ch=81$ is used for processing each subband LL, LH, HL, HH. In Figure \ref{fig:depen}, $ch=32$ is used. Our codes are available on github at \cite{berkgithub}.



\ifCLASSOPTIONcaptionsoff
  \newpage
\fi



%
\bibliographystyle{IEEEtran}
\bibliography{thesis.bib}{}


%

\vspace{-1.1cm}

\begin{IEEEbiographynophoto}{Uğur Berk Şahin}
received the B.S. degree and the M.S degree in electrical and electronics engineering from Middle East Technical University, Ankara, Turkey, in 2019 and 2022, respectively. He is currently working as computer vision engineer in Aselsan Inc. His research interests are image compression, object detection and visual intertial odometry.
\end{IEEEbiographynophoto}
\vspace{-1.1cm}
\begin{IEEEbiographynophoto}{Fatih Kamışlı}
received the B.S. degree in electrical and electronics engineering from Middle East Technical University, Ankara, Turkey, in 2003, and the M.S. and Ph.D. degrees in electrical engineering and computer science from the Massachusetts Institute of Technology, Cambridge, MA, USA, in 2006 and 2010, respectively. He is currently an Associate Professor in the Electrical and Electronics Engineering Department at the Middle East Technical University. His  research interests include image and video compression.
\end{IEEEbiographynophoto}





\end{document}